# Time-dependent Interactions in Tunnelling Dynamics

Luca Nanni

luca.nanni@edu.unife.it

**Abstract** In this paper, the tunnelling of a particle through a potential barrier is investigated in the presence of a time-dependent perturbation. The latter is attributed to the process of the energy measurement of the scattered particle. The time-dependent Schrödinger equation of the model is exactly solved. The calculation of the probability density inside the barrier proves that the tunnelling dynamics is determined not only by the transmitted and reflected waves but also by their interference. Furthermore, the interference term is time-dependent and contribute to the scattering process duration. The tunnelling time is calculated as the time to stop the flow of probability density inside the barrier. This is the minimum duration of the measurement process before detecting the particle beyond the barrier. Based on this, a new method of estimating the tunnelling time by energy experimental measuring is proposed.

**1. Introduction**

Tunnelling is a phenomenon peculiar to the quantum world that has been widely applied in physics, chemistry, and on which is based the functioning of many technological devices, such as diodes, superconducting quantum interference devices, quantum antennas and superconducting qubits for quantum computers [1-8]. The quantum tunnelling problem can be addressed using two distinct approaches. The first is the time-independent approach relying on the principle of conservation of energy [9-12]. In this case, the energy of the particle scattered through the potential barrier is equal to its initial energy. The second approach is the time-dependent one, which is based upon the perturbation theory [13-16]. In this case, the potential barrier is considered as a perturbation and the energy of the scattered particle is spreads within a small range. However, an accurate description of the tunnelling process cannot be made using just one of these approaches. For instance, the measuring of the energy of a particle scattered by a rectangular potential barrier, whose value does not depend on time, can be performed only when the tunnelling has ended, i.e. after a time $\tau$. This means that, according to the Heisenberg uncertainty principle, the uncertainty affecting the energy of the scattered particle is $\delta E \geq \hbar/2\tau$. Therefore, is impossible to state whether the tunnelling is stationary or non-stationary, especially when the configuration of the barrier is such as to involve very short tunnelling times. In these cases, the uncertainty affecting the particle energy can be of the order of the height of the barrier.

In this work we study the tunnelling process of a particle through a potential barrier in presence of a time-dependent interaction, due to a measurement process involved in the tunnelling. The tunnelling dynamics of this model is investigated by the time-dependent

Schrödinger equation (TDSE), which in the non-relativistic framework represents the most suitable tool for dealing with similar problems. The TDSE is solved exactly and the probability density of the particle within the barrier is calculated. Finally, the tunnelling time is calculated both as the time required to stop the flow of the probability current and by the transfer matrix method. We believe that these approaches are the most appropriate for a non-stationary tunnelling problem, as is the one being investigated. In fact, the tunnelling times defined in literature (dwell time, phase time, Larmor time, complex time) [17] mainly relate to stationary processes and are calculated as average values obtained by integrating on all the scattering channels. In this case, the flow of probability density reaches a steady-state and is maintained for the entire duration of the scattering. But in a measurement process, this does not occur, and hence we need to change the way to calculate the tunnelling time. The configuration of the potential barrier determines the tunnelling time and allows to estimate a priori the uncertainty affecting the measurement of the energy of the scattered particle.

## 2. Exact Solution of TDSE of the Model

Let us consider a particle of mass $m$ moving along $x$ axis. In the region $0 \leq x \leq L$ a potential barrier $U(t,x)$ is present:

$$U(t,x) = U(x) + U(t) \quad 0 \leq x \leq L, \qquad (1)$$

where $U(x)$ is the potential for each point of the barrier and $U(t)$ is a time-dependent interaction potential. We do not make any assumptions about the geometry of the barrier. In the regions $x < 0$ and $x > L$ the potential is everywhere zero. We are interested investigating the dynamics of the particle within the barrier. The TDSE for this model is:

$$i\hbar \frac{\partial}{\partial t}\psi(t,x) = \left[-\frac{\hbar^2}{2m}\frac{\partial^2}{\partial x^2} + U(t,x)\right]\psi(t,x). \qquad (2)$$

Substituting Eq. (1) in Eq. (2) is obtained:

$$i\hbar \left[\frac{\partial}{\partial t} - U(t)\right]\psi(t,x) = \left[-\frac{\hbar^2}{2m}\frac{\partial^2}{\partial x^2} + U(x)\right]\psi(t,x). \qquad (3)$$

Since we supposed the spatial and temporal contributions to the potential are independent, the wavefunction $\psi(t,x)$ can be factorized as follows:

$$\psi(t,x) = \varphi(x)\vartheta(t). \qquad (4)$$

Using Eq. (4), the Eq. (3) is split into two separate equations:

$$\begin{cases} i\hbar \left[\frac{\partial}{\partial t} - U(t)\right]\vartheta(t) = E\vartheta(t) \\ \left[-\frac{\hbar^2}{2m}\frac{\partial^2}{\partial x^2} + U(x)\right]\varphi(x) = E\varphi(x) \end{cases}, \qquad (5)$$

where $E$ is the particle energy. The first of Eq. (5) admits the following general solution:

$$\vartheta(t) = \varepsilon e^{\left[-\frac{iEt}{\hbar} - \frac{i}{\hbar}\int_0^t U(t)dt\right]}, \qquad (6)$$

where $\varepsilon$ is an arbitrary coefficient. The solution of the second of Eq. (5) can be represented as the linear combination between the transmitted component of the incident wave and the component reflected by the right side of the barrier, denoted respectively by $\varphi_T(x)$ and $\varphi_R(x)$. These two components are evanescent waves characterized by an imaginary wave vector $\chi = \pm\sqrt{2m(E-U)}/\hbar$ and they do not depend on the time being localized waves [18-20]. Therefore, in the formula that yields $\chi$, only the term $U(x)$ must be considered. For convenience we write $\varphi_T(x)$ and $\varphi_R(x)$ as follows:

$$\varphi_T(x) = \alpha f(x) \quad ; \quad \varphi_R(x) = \beta g(x), \tag{7}$$

where $\alpha$ and $\beta$ are numerical coefficient and $f(x)$ and $g(x)$ are evanescent waves whose explicit form is:

$$f(x) = e^{-|\chi|x/\hbar} \quad ; \quad g(x) = e^{|\chi|x/\hbar}. \tag{8}$$

It must be understood that the value of the imaginary wave vector $\chi$ can vary within the range $0 \le x \le L$, depending on the geometry of the barrier. The coefficients $\alpha$ and $\beta$ are obtained imposing the following boundary conditions:

$$\psi_i(x)|_{x=0} = \alpha f(x)|_{x=0} \quad ; \quad \alpha f(x)|_{x=L} = \beta g(x)|_{x=L}, \tag{9}$$

where $\psi_i(x)$ is the spatial part of the incident wave function, which has the form of a plane wave with real wave vector $K = \sqrt{2mE}/\hbar$ [2]. Therefore, the general solution of Eq. (3) in the range $0 \le x \le L$ is:

$$\psi(t,x) = c_1 \varphi_T(x) e^{\left[-\frac{i}{\hbar}\int_0^t U(t)dt\right]} e^{\left[-\frac{iE_k t}{\hbar}\right]} + c_2 \varphi_R(x) e^{\left[-\frac{i}{\hbar}\int_0^t U(t)dt\right]} e^{\left[-\frac{iE_j t}{\hbar}\right]}, \tag{10}$$

where $c_1$ and $c_2$ are arbitrary coefficients that are obtained normalizing the wave function. The terms $E_k$ and $E_j$ are two possible energy values that the particle can assume inside the potential barrier at different spacetime points. In fact, Eq. (10) is compatible with a simultaneous measurement of the particle energy at the two sides of the barrier performed in the time interval $\Delta t = (t - 0)$. As prescribed by quantum measurement theory, driven by Heisenberg's uncertainty principle, the two energy values have expected to be different [21].

Let us now substituting Eq. (10) in the Eq. (2) obtaining:

$$i\hbar\dot{a}_k(t)\varphi_T(x) + i\hbar\dot{a}_j(t)\varphi_R(x)e^{i\omega_{kj}t} = a_k(t)U(t)\varphi_T(x) + a_j(t)U(t)\varphi_R(x)e^{i\omega_{kj}t}, \tag{11}$$

where:

$$\begin{cases} a_{k(j)}(t) = \gamma_{k(j)} e^{\left[-\frac{i}{\hbar}\int_0^t U(t)dt\right]} \\ \omega_{kj} = \omega_k - \omega_j = (E_k - E_j)/\hbar \end{cases}. \tag{12}$$

In Eq. (12) $\gamma_{k(j)}$ are arbitrary coefficients. The time-dependent coefficients $a_{k(j)}(t)$ are obtained multiplying both sides of Eq. (11) by $\varphi_T^*(x)$ first and by $\varphi_R^*(x)$ then, and integrating in the range $0 \le x \le L$:

$$i\hbar\dot{a}_k(t)\int_{x_j}^{x_k}\varphi_T^*(x)\varphi_T(x)dx + i\hbar\dot{a}_j(t)e^{i\omega_{kj}t}\int_{x_j}^{x_k}\varphi_T^*(x)\varphi_R(x)dx = a_k(t)U(t)\int_{x_j}^{x_k}\varphi_T^*(x)\varphi_T(x)dx + a_j(t)U(t)e^{i\omega_{kj}t}\int_{x_j}^{x_k}\varphi_T^*(x)\varphi_R(x)dx, \tag{13}$$

$$i\hbar\dot{a}_k(t)\int_{x_j}^{x_k}\varphi_R^*(x)\varphi_T(x)dx + i\hbar\dot{a}_j(t)e^{i\omega_{kj}t}\int_{x_j}^{x_k}\varphi_R^*(x)\varphi_R(x)dx =$$
$$a_k(t)U(t)\int_{x_j}^{x_k}\varphi_R^*(x)\varphi_T(x)dx + a_j(t)U(t)e^{i\omega_{kj}t}\int_{x_j}^{x_k}\varphi_R^*(x)\varphi_R(x)dx. \quad (14)$$

From Eq. (13) and Eq. (14) is obtained:

$$\begin{cases} i\hbar\dot{a}_k(t)X_{jk} + i\hbar\dot{a}_j(t)X_{jj}e^{i\omega_{kj}t} = a_k(t)Y_{kj} + a_j(t)Y_{jj}e^{i\omega_{kj}t} \\ i\hbar\dot{a}_k(t)X_{kk} + i\hbar\dot{a}_j(t)X_{kj}e^{i\omega_{kj}t} = a_k(t)Y_{kk} + a_j(t)Y_{jk}e^{i\omega_{kj}t}. \end{cases} \quad (15)$$

In Eq. (15) $X_{jk}$ and $Y_{jk}$ are respectively the components of the overlapping and transition matrices, given by:

$$\begin{cases} X_{kj} = \int_{x_j}^{x_k}\varphi_T^*(x)\varphi_R(x)dx \\ Y_{jk} = \int_{x_j}^{x_k}\varphi_T^*(x)U(t)\varphi_R(x)dx \end{cases} \quad (16)$$

where $x_k, x_j \in [0, L]$. From Eq. (16) one sees that $Y_{kk} = Y_{jj} = 0$. As one can be guessed, the theory being formulating is very similar to the spectroscopic one and suggests that tunnelling can be interpreted as transition from a state $\varphi_T(x)$ to a state $\varphi_T(x)$ induced by the potential $U(t)$. The Eq. (16) is a system of two linear differential equations where unknown functions are $a_k(t)$ and $a_j(t)$. Separating these unknown functions and integrating respect the time is obtained:

$$\begin{cases} a_k(t) = e^{i\omega_0 t} \\ a_k(t) = -i\omega_0 \dfrac{X_{kk}}{X_{kj}} e^{i\omega_0 t} \dfrac{sin(\omega t/2)}{(\omega/2)}, \end{cases} \quad (17)$$

where:

$$\omega_0 = \frac{X_{kj}Y_{jk}}{(X_{kk}X_{jj}-X_{kj}X_{jk})} \quad and \quad \omega = \omega_{jk} - \omega_0. \quad (18)$$

With Eq. (17) the TDSE is thus exactly solved.

## 3. Tunnelling Probability Density

In this section, the probability density inside the barrier is calculated. Using Eq. (10) together with Eq. (17) we obtain:

$$\rho(t,x) = \psi^*(t,x)\psi(t,x) = \varphi_T^*(x)\varphi_T(x) +$$
$$\omega_0^2 \frac{|X_{kk}|^2}{|X_{kj}|^2}\varphi_R^*(x)\varphi_R(x)\frac{sin^2(\omega t/2)}{(\omega/2)^2} + 2\omega_0 \frac{X_{kk}}{X_{kj}}\varphi_T^*(x)\varphi_R(x)\frac{sin^2(\omega t/2)}{(\omega/2)}. \quad (19)$$

The third term in the right-hand-side of Eq. (19) represents the interference between the transmitted and reflected waves. Let us consider, for simplicity, the case of a rectangular barrier where $U = U_0 \; \forall x \in [0, L]$. Then Eq. (19) becomes:

$$\rho(t,x) = |\alpha|^2 e^{-2|\chi|x} + |\beta|^2 \omega_0^2 \frac{|X_{kk}|^2}{|X_{kj}|^2} e^{2|\chi|x} \frac{sin^2(\omega t/2)}{(\omega/2)^2} +$$
$$2\omega_0 \frac{X_{kk}}{X_{kj}} \alpha^* \beta \frac{sin^2(\omega t/2)}{(\omega/2)}, \quad (20)$$

where in writing Eq. (20) have been used Eq. (7) and Eq. (8). Moreover, using the second of boundary conditions given by Eq. (9) is proved that:

$$|\beta|^2 \omega_0^2 \frac{|X_{kk}|^2}{|X_{kj}|^2} = 4 \frac{|\alpha|^2 \hbar^2 |\chi|^4}{m^2} e^{-2|\chi|x}. \tag{21}$$

Substituting Eq. (21) in Eq. (20) we obtain:

$$\rho(t,x) = |\alpha|^2 e^{-2|\chi|x} + 4 \frac{|\alpha|^2 \hbar^2 |\chi|^4}{m^2} \frac{\sin^2(\omega t/2)}{(\omega/2)^2} + 2\omega_0 \frac{X_{kk}}{X_{kj}} \alpha^* \beta \frac{\sin^2(\omega t/2)}{(\omega/2)}, \tag{22}$$

As expected, when $t = 0$, i.e. when the measurement process has not yet started, the probability density tends asymptotically to zero as $x$ tends to L. But as soon as $t \neq 0$ then the probability density oscillates in every point inside the barrier between a maximum and a minimum given by:

$$\begin{cases} \rho_{min}(t,x) = |\alpha|^2 e^{-2|\chi|x} \\ \rho_{max}(t,x) = |\alpha|^2 e^{-2|\chi|x} + 4 \frac{|\alpha|^2 \hbar^2 |\chi|^4}{m^2 (\omega/2)^2} + 2\omega_0 \frac{X_{kk}}{(\omega/2) X_{kj}} \alpha^* \beta \end{cases} \tag{23}$$

Therefore, once set a point $x'$ inside the barrier, the probability density is spread over time in a range of values which is wider the smaller the energy difference $(E_k - E_j)$, the higher the potential barrier and the greater the interference between transmitted and reflected wave. Eq. (23) represents the tool by which is possible to choose which initial parameters to modify to modulate the performance of an electronic device that based on quantum tunnelling.

In the case the form of the potential $U(x)$ is different from the rectangular one, the procedure discussed above does not change except for possible mathematical complications in the calculation of the transition integrals.

## 4. Tunnelling Time

The second step of this study is to estimate the tunnelling time. In fact, the time required to perform the measurement of the energy of the scattered particle must be at least equal to the time needed to complete the tunnelling process. Tunnelling time is one of the most debated and controversial topics in quantum mechanics, both in the theoretical and experimental framework [22-26]. There are different definitions of tunnelling time and all of them present weaknesses [17]. Moreover, these definitions refer to stationary processes, while in this study we are addressing a non-stationary tunnelling problem. In the case being investigated, related to a non-stationary problem, the tunnelling time is calculated as the time needed to stop the flow of probability density inside the barrier. The latter is given by the following derivative [27]:

$$\frac{\partial \rho(t,x)}{\partial x} = -2|\alpha|^2 |\chi| e^{-2|\chi|x} \left[1 - 4 \frac{\hbar^2 |\chi|^4}{m^2} \frac{\sin^2(\omega t/2)}{(\omega/2)^2}\right]. \tag{24}$$

Eq. (24) vanishes when the term in square brackets equals zero. This implies that:

$$t = \tau_0 = \frac{2}{\omega_0} arcsin\left(\frac{m\omega_0}{4\hbar |\chi|^2}\right). \tag{25}$$

Therefore, Eq. (25) yields the tunnelling time. This time depends on the energy difference and on the height of the potential barrier, but not on the barrier length. Therefore, Eq. (25) implicitly predicts the Hartman effect [28] without having to apply any mathematical approximation. This is a relevant result that, to the best of our knowledge, is not mentioned in literature. If the argument of the arcsine function is small enough, Eq. (25) can be simplified obtaining:

$$\tau_0 = \frac{m\omega_0}{2\hbar|\chi|^2} = \frac{\hbar}{4(U_0-E)}. \qquad (26)$$

Eq. (26) shows more clearly that the tunnelling time is shorter the higher the barrier is, and this is exactly what is achieved in tunnelling diodes, where the potential barrier between the two semiconductors is high and narrow.

Another approach that can be used to calculate the tunnelling time is the one based on the transfer matrix method. The latter provides an important tool for investigating bound and scattering states in quantum structures. It is mainly used to solve the one-dimensional Schrodinger or effective mass equation, e.g., to obtain the quantized energies in quantum well heterostructures and metal-oxide-semiconductor structures [29] or the transmission coefficient of potential barriers [30]. Therefore, this method is suitable to calculate the tunnelling time also for a non-stationary case. For a one dimensional scattering problem, like the one being investigated, the tunnelling time is given by [31]:

$$\tau_0 = \frac{1}{|\alpha|} \int_0^L \sqrt{\frac{m}{E_{inc.}-E(x)}} dx. \qquad (27)$$

where $E_{inc.}$ is the energy of the incident particle at $x = 0$ and $E(x)$ is the energy of the particle in a given point $x$ inside the barrier. From Eq. (27) is clear that the tunnelling time is shorter the greater the transmission coefficient $|\alpha|$ and the greater the dispersion of the tunnelling-particle energy $(E_{inc.} - E(x))$ induced by the measurement process. In the tunnelling time of Eq. (26), the energy dispersion effect of the tunnelling particle is implicit in the term $\omega_0$, through the integral given by the second of Eq. (16). In Eq. (27), on the other hand, the dependence on the barrier height is contained in the transmission coefficient $|\alpha|$, which is proportional to the imaginary wave vector $\chi = \pm\sqrt{2m(E-U)}/\hbar$ whose explicit form contains the potential $U$. Therefore, the two approaches used to calculate the tunnelling time lead to the same conclusions. However, the formula of Eq. (27) highlights in a more direct way the dependence of F on the perturbation due by the measurement process, anticipating what we will discuss shortly.

Let us now return to the problem of measuring the energy of the particle exiting the barrier. As mentioned at the beginning of this section, the time taken to perform this measurement cannot be less than the tunnelling time. This is reflected in the error affecting the energy, which is greater the shorter the tunnelling time, in accordance with the uncertainty principle $\delta E \geq \hbar/2\tau_0$. In absence of the time-dependent interaction due to the measurement process, the energy of the scattered particle is equal to that of the incident particle. Therefore, we can infer that the measurement error on the energy is given by $\delta E = (E_{inc.} - E_{scatt.}^{meas.})$, where $E_{scatt.}^{meas.}$ is the measured energy of the scattered particle. We have thus obtained an indirect way of measuring the tunnelling time:

$$\tau_0^{meas.} \geq \hbar/2(E_{inc.} - E_{scatt.}^{meas.}). \tag{28}$$

Eq. (28) thus allows calculating the minimum tunnelling time for any device based on quantum tunnelling, regardless the form of the potential barrier. Knowing a priori the particle mass, its initial energy and the height of the barrier, from Eq. (28) is possible backwards to calculate the angular frequency $\omega_0$ and therefore, through Eq. (18), to have information on the nature of the scattering dynamics inside the barrier. This is a new approach to study the tunnelling processes and represents the novelty of this work. The problem remains that of performing a weak measurement to not excessively disturb the quantum system. Our approach to calculating tunnelling time is reminiscent of Steinberg's, in which ultracold rubidium atoms are propelled gently through a barrier induced by a light beam [32]. In the experiment is measured the change of the spin orientation of the atoms when they exit the barrier. The amount of this change is proportional to the time spent by the atoms inside the barrier. In the theory we propose, the same experiment should be performed measuring the energy change of the particle exiting the barrier.

**5. Discussion**

Understanding the dynamics governing quantum tunnelling is of main importance to improve or design new tunnelling-based devices. This means performing experimental measurements which, as is well known in the framework of quantum mechanics, involve interactions with the system. Hence the need to investigate the tunnelling process of a particle through a potential barrier, of any shape, in the presence of a time-dependent perturbation. In this study, the time-dependent Schrodinger equation of the particle inside the potential barrier have been solved exactly, with the aim of calculating the probability density and obtaining information on the possible processes that take place within the barrier. As expected, the probability density is given not only by the contribution of the transmitted and reflected waves, but also by their interference. What emerges, however, is that the latter contribution is time-dependent and represents the dynamics by which the measurement process perturbs the tunnelling. This interaction contributes to the tunnelling time which, in the framework being studied, can be interpreted as the minimum time to measure the energy of the scattered particle. Starting from this assumption and invoking the Heisenberg uncertainty principle, is possible estimating the tunnelling time from the experimental measurement of the particle energy after tunnelling, assuming that the uncertainty is given by the difference between the initial energy and the measured energy. This procedure allows to deal with tunnelling time regardless of the possible definitions proposed in literature and to obtain information on the dynamics of the processes that take place inside the barrier.